          [2001/04/25 1.1 (PWD)]
\documentclass[a4paper,twocolumn]{esapub2005} % European paper
\usepackage{amsmath}
\usepackage{txfonts}
\usepackage{graphicx}
\usepackage{amsfonts}
\usepackage{times}
\usepackage{natbib}

\title{12 Bootis: the need for asteroseismic constraints}
\author[1]{A. Miglio}
\author[1]{J. Montalb\'an}
\author[2]{C. Maceroni}
\author[3]{J. De Ridder}
\author[2]{F. D'Antona}

\affil[1]{Institut d'Astrophysique et G\'eophysique de l'Universit\'e de Li\`ege, Belgium}
\affil[2]{INAF, Osservatorio Astronomico di Roma, Italy}
\affil[3]{Instituut voor Sterrenkunde, Katholieke Universiteit Leuven, Belgium}

\newcommand{\BV}{Brunt-V\"ais\"al\"a }
\newcommand{\msol}{\mbox{${\rm M}_{\odot}$}}

\newcommand{\lsol}{\mbox{${\rm L}_{\odot}$}}
\newcommand{\teff}{\mbox{${T}_{\rm eff}$}}
\begin{document}

%\keywords{; ; }

\maketitle

\begin{abstract}
12 Bootis is a double-lined spectroscopic binary whose orbit has been resolved by interferometry. We present a detailed modelling of the system and show that the available observational constraints can be reproduced by models at different evolutionary stages, depending on the details of extra-mixing processes acting in the central regions. In order to discriminate among these theoretical scenarios, additional and independent observational constraints are needed: we show that these could be provided by solar-like oscillations, that are expected to be excited in both system components.
\end{abstract}

\section{Observational constraints}
The physical parameters of the system have been precisely determined by \citep{Boden05} combining interferometric and spectroscopic observations. A further improvement in the determination of the spectroscopic orbit by \citep{Tomkin06}, combined with interferometric data, confirms the values of the masses derived by \citep{Boden05} and reduces the associated error bars. The observational constraints we assume in our modelling are taken from Table 5 of \citep{Boden05} that, for the sake of convenience, we report here in Table \ref{tab:obs}. As reviewed in \citep{Boden05} spectroscopic studies of 12 Bootis in the literature (\citep{Lebre99}; \citep{Balachandran90}) indicate a metallicity within 0.1 dex of the solar value.

\section{Modelling 12 Bootis}
\begin{table}
  \begin{center}
    \caption{Observational constraints adopted in the modelling.}\vspace{1em}
    \renewcommand{\arraystretch}{1.2}
\begin{tabular}[h]{lcc}
  \hline
  & A & B \\
\hline
M/\msol & 1.4160$\pm$0.0049& 1.3740$\pm$0.0045\\
\teff~(K)& 6130$\pm$100 & 6230$\pm$150 \\
L/\lsol & 7.76$\pm$0.35 & 4.69$\pm$0.74\\
$[\rm Fe/H]$ & 0.0$\pm$0.1 & 0.0$\pm$0.1 \\
 \hline
 \end{tabular}
    \label{tab:obs}
  \end{center}
\end{table}

\begin{figure}
\centering
\includegraphics[width=\linewidth]{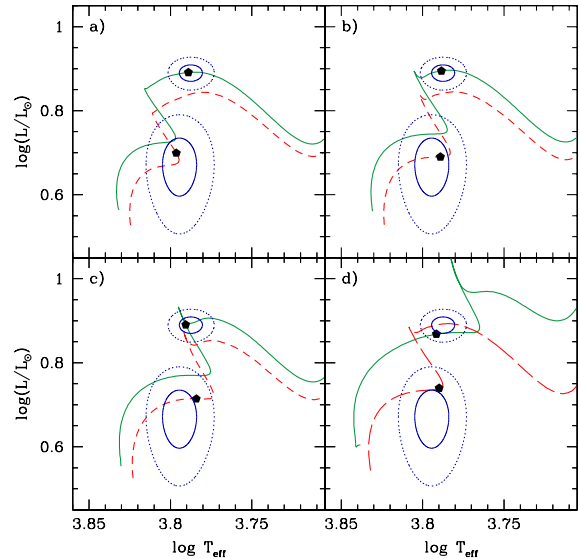}
\caption{ HR diagram showing the solutions found with different values of the overshooting parameter: {\it a)} $\alpha_{\rm OV} = 0.06$, {\it b)} $\alpha_{\rm OV} = 0.15$, {\it c)} $\alpha_{\rm OV} = 0.23$, and {\it d)}  $\alpha_{\rm OV,A} = 0.37$ and $\alpha_{\rm OV,B} = 0.15$. The parameters of the models are described in Table \ref{tab:results}. Also shown in continuous and dotted lines are, respectively, the 1- and 2-$\sigma$ error ellipses in $T_{\rm eff}$ and luminosity.}\label{fig:hrall}
\end{figure}
We model the binary system considering as free parameters the initial chemical composition (described by the the hydrogen and heavy-elements mass fraction, X and Z), the amount of overshooting, the age and the masses of the components. These parameters are changed in order to reproduce the observational constraints presented in the previous section, as well as  the additional constraint of  same age and initial chemical composition for both components. A standard $\chi^2$ is used as a goodness-of-fit measurement so that each observable is taken into account along with its uncertainty.
The stellar models are computed using CLES (Code Liégeois d'Evolution  Stellaire). The standard physics used is: OPAL01 EOS \citep{Rogers02}; OPAL96 \citep{Iglesias96} opacity tables  complemented with \citep{Alexander94} at $T < 10000\; K$. Nuclear reaction rates from \citep{Caughlan88} updated with the recent measurement of the $^{14}{\rm N}(p,{\gamma})^{15}{\rm O}$ reaction rate \citep{Formicola04}; standard solar mixture \citep{grevesse93}, and  atmospheric boundary conditions at $T=T_{\rm eff}$ given by \citep{Kurucz98}. Chemical mixing in the formal convective region (Schwarzschild criterion) and in the overshooting layer is treated as instantaneous, the thickness of the overshooting layer $\Lambda_{\rm OV}$ is expressed in terms of a parameter $\alpha_{\rm OV}$, so that $\Lambda_{\rm OV} =\alpha_{\rm OV}×\min( r_{cc}, H_p(r_{cc}) )$, where  $r_{cc}$ is the radius of the convective core.

As noticed  in \citep{Boden05} the result of the fit is highly dependent on the model details. We model the system considering different values of overshooting (the same for both components) and find solutions for $\alpha_{\rm OV} \ge 0.06$. The evolutionary state of the solution depends on $\alpha_{\rm OV}$ , as described below (see also Table \ref{tab:results}).
For the lowest  values of overshooting the situation is the one  described in the upper-left panel of Fig. \ref{fig:hrall}, that is: the primary component close to the maximum luminosity in the shell--hydrogen burning phase, whereas the secondary, with a Xc=0.06, is at the end of its central hydrogen-burning phase.
 As  $\alpha_{\rm OV}$ increases, the evolutionary state of A component approaches the TAMS. With $\alpha_{\rm OV} \simeq 0.15$ (upper-right panel of Fig. \ref{fig:hrall}) the primary is burning hydrogen in a thick shell while the component B is well on the main sequence. A  further increase of the overshooting parameter $\alpha_{\rm OV} \gtrsim 0.2$ places the primary in the rapid overall contraction phase (Fig. \ref{fig:hrall}, lower-left panel) whereas, for even larger values of $\alpha_{\rm OV}$, $L_{\rm B}$ increases out of the error ellipse and therefore worsens the fit.

If we seek a solution where both components are on the Main-Sequence, then the main obstacle is to reproduce the observed luminosity ratio $L_{\rm A}/L_{\rm B} \sim 1.65$. In fact, as both components have the same age and  initial chemical composition, and if the same amount of extra-mixing in the core is considered, the luminosity ratio in the main-sequence is determined  by the difference of mass between the components ($q\sim0.97$), and does not significantly depend on the choice of the other parameters in the modelling.
Regardless of the choice of the parameter set ( $X,Z,\alpha_{\rm OV}$), all the models  computed with $\alpha_{\rm OV,A}=\alpha_{\rm OV,B}$ provide $(L_{\rm A}/L_{\rm B})_{\rm MS} \simeq 1.15$. 
A higher value of the luminosity ratio can be reached if  $\alpha_{\rm OV,A} > \alpha_{\rm OV,B}$. Therefore, the only possibility we find to place both components in their MS phase is to assume a different effect of  mixing processes in component A and B, as in the model 
plotted in the lower-right panel of Fig.~\ref{fig:hrall} (model $d$ in Table \ref{tab:results}).
% (lower-right panel of Fig.~\ref{fig:hrall})}.
%risultati fit
\begin{table}
  \begin{center}
    \caption{Parameters of the models presented in Fig. \ref{fig:hrall}.}\vspace{1em}
    \renewcommand{\arraystretch}{1.2}
\label{tab:results}
\begin{tabular}{lcccc}
\hline
&  $\alpha_{\rm OV}$ & $Y_0$ & $Z_0$ & Age (Gyr) \\
\hline
a  & 0.06        & 0.268 & 0.0195 & 3.08\\
b  & 0.15        & 0.267 & 0.0198 & 3.30\\
c  & 0.23        & 0.267 & 0.0200 & 3.58\\
d  & 0.37, 0.15  & 0.281 & 0.0186 & 3.24\\
\hline
 \end{tabular}
  \end{center}
\end{table}
\subsection{Solar-like oscillations}\label{sec:grf}
Solar-like oscillations are expected to be excited in both components of the system. As the primary is expected to dominate the observed light spectrum, due to its larger luminosity, we here discuss the seismic properties solely of 12 Boo A. 

As the components of 12 Bootis are massive enough to develop a convective core during the main sequence, the combined action of nuclear burning and convective mixing is responsible for the development of a steep chemical composition gradient at the boundary of the convective core.
As a star leaves the main-sequence, the increasing central condensation, together with the steep chemical composition gradient in the central regions, leads to a large increase of the buoyancy frequency and therefore of the frequencies of gravity modes. The latter interact with p-modes and affect the properties of non-radial solar-like oscillations by avoided crossings (see e.g. \citep{Osaki75}, \citep{Aizenman77}).
As already investigated in the case of $\eta$ Bootis \citep[see e.g.][]{Dimauro04}, the frequencies of modes undergoing an {\it avoided-crossing} are direct probes of the central structure of intermediate mass stars and, therefore, of their evolutionary status.
\begin{figure}
\centering
\includegraphics[width=\linewidth]{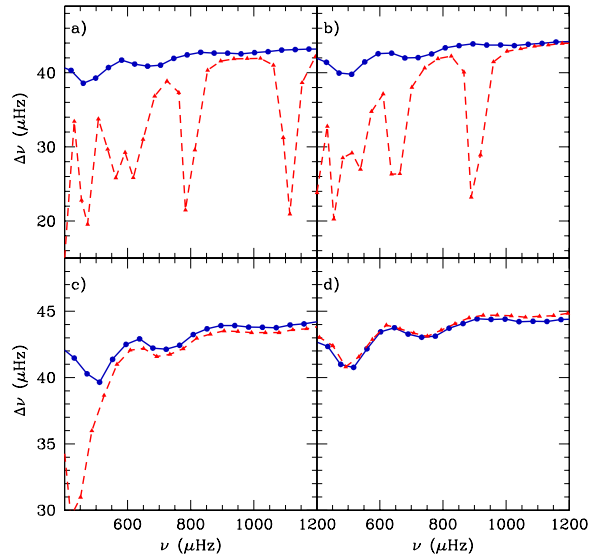}
\caption{ Large frequency separation as a function of the frequency for the models described in Fig. \ref{fig:hrall} and \ref{fig:nall}. Solid and dashed lines connect, respectively, $\Delta\nu$ for radial and $\ell=1$ modes. The frequencies of $\ell=1$ modes provide a test to discriminate among the different theoretical scenarios. \label{fig:large}}
\end{figure}

In Fig. \ref{fig:large} we compare the large frequency separation $\Delta\nu$ for radial and $\ell=1$ modes computed for the models shown in Fig. \ref{fig:hrall}.
Even though radial modes do not give information on the evolutionary status of the star, since they are mainly determined by the stellar mass and radius, $\ell=1$ modes allow a clear discrimination among the scenarios.
%Though radial modes do not give information on the evolutionary status, $\ell=1$ modes allow a clear discrimination among the scenarios.
In the domain of solar-like oscillations, whereas modes of mixed p and g character are not present in the model on the main sequence (Fig. \ref{fig:large}, {\it panel d}), they do affect, at relatively low frequencies, the model in the overall contraction phase ({\it panel c}) and break the regular frequency spacing in more evolved models ({\it panel a and b}).

\begin{figure}
\centering
\includegraphics[width=\linewidth]{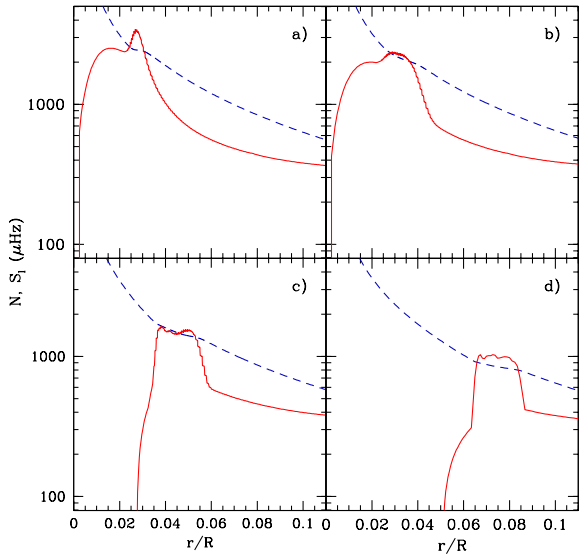}
\caption{Propagation diagram in the core of the models in Fig. \ref{fig:hrall} and \ref{fig:large}. The different evolutionary state reflect the different value of $\alpha_{\rm OV}$ adopted the models. The solid line represents the \BV frequency, the dashed line the Lamb frequency for $\ell=1$.\label{fig:nall}}
\end{figure}

In the models considered, the radical variation of the properties of $\ell=1$ modes derives from a considerable change of the \BV frequency near the center of the star (see Fig. \ref{fig:nall}) as hydrogen is exhausted throughout the convective core and the star undergoes an overall contraction.

\section{Discussion}
The binary system 12 Bootis represents as an ideal candidate to test extra-mixing processes in intermediate-mass stars. The combination of the already available precise observational constraints with solar-like oscillations would allow a reliable determination of the extension of the convective core during the main sequence and of the properties of the central region of the star.
A more detailed modelling of the system (including the effects of diffusive overshooting \citep{Ventura98} and microscopic diffusion \citep{Thoul94}) will be presented in a forthcoming paper.

\section*{Acknowledgments}
A.M. and J.M  acknowledge financial support from the Prodex 8 COROT (C90199). A.M. is also thankful to HELAS and FNRS for financial support.

% The following bibliography was produced with
%   \bibliographystyle{aa}
%   \bibliography{esapub}
% The results are inserted directly here to simplify
% the demonstration.

%\begin{thebibliography}
   \bibliographystyle{unsrt}
   \bibliography{andrea}
%\bibitem[Allen(1973)]{allen73}
%Allen C., 1973, Astrophysical Quantities, Athlone Press
%\bibitem[Nobody et~al.(1997)]{nobody97}
%Nobody B., Somebody G., Who M.E., et~al., 1997, ApJ 331, 902
%\bibitem[Smith \& Jones(1996)]{smith96}
%Smith A., Jones B., 1996, A\&A 555, 999

%\end{thebibliography}
\end{document}